**2D graphitic-like gallium nitride and structural selectivity in confinement at graphene/SiC interface**

*Gianfranco Sfuncia,[a] Giuseppe Nicotra,[a]\* Filippo Giannazzo,[a] Béla Pécz,[b] Gueorgui Kostov Gueorguiev,[c] and Anelia Kakanakova-Georgieva,[c]\**

[a] Consiglio Nazionale delle Ricerche, Istituto per la Microelettronica e Microsistemi, Strada VIII, n. 5, Zona Industriale, I-95121, Catania, Italy

[b] Centre for Energy Research, Institute of Technical Physics and Materials Science, Konkoly-Thege M. út 29-33., Budapest, 1121, Hungary

[c] Department of Physics, Chemistry and Biology (IFM), Linköping University, 581 83 Linköping, Sweden

\* E-mail: anelia.kakanakova@liu.se

\* E-mail: giuseppe.nicotra@cnr.it

**Keywords**: 2D GaN, 2D $Ga_2O_3$, MOCVD, C-AFM, S/TEM, EELS

**Abstract**: Predictive first-principles calculations suggest graphitic-like GaN to be theoretically possible. Thus far, it has not been experimentally reported. We report on GaN monolayer in a buckled geometry obtained in confinement at graphene/SiC interface by metalorganic chemical vapor deposition (MOCVD). Conductive atomic force microscopy (C-AFM) has been employed to probe vertical current injection through the graphene/SiC interface and to establish the uniformity of the intercalated areas. Scanning transmission electron microscopy (S/TEM) has been employed for atomic resolution imaging and spectroscopy. Discontinuity in the anticipated stacking sequence of graphitic-like GaN monolayers has been exposed and reasoned as a case of simultaneous formation of Ga-N and Ga-O bonds. The formation of Ga-O bonds acquires importance in instigating chemical-species-specific structural selectivity in confinement at atom-size scale.

## 1. Introduction

Graphitic-like group III nitrides - including experimentally demonstrated AlN [1] and InN [2] by metalorganic chemical vapor deposition (MOCVD) – have become a paradigm for a class of conceptual 2D materials. A few honeycomb monolayers - each with planar trigonal $sp^2$ coordination between group III and nitrogen atoms – stacked in an ultrathin graphitic-like structure is considered as the ultimate scaling down in the vertical (c-axis) direction for AlN, GaN and InN. Instead, their 3D bulk is represented by a wurtzite structure with tetrahedral $sp^3$ coordination between the atoms. From an alternative perspective, the ultrathin few-layer films of graphitic-like group III nitrides have been perceived as acting as precursors to respective thin films with wurtzite structure [3]. The relaxation of the ultrathin graphitic-like films to the wurtzite structure, which becomes relevant at accumulation of monolayers beyond a certain number depending on the material composition [3], and the underlying charge transfer mechanism can be associated with increasing electric dipole moment component in the perpendicular direction to the atomic plane [4].

The honeycomb monolayer structure of group III nitrides, as well as other group III-V semiconductors, has been subjected to substantial exploration by first-principles calculations based on density functional theory [5-7]. Predictive first-principles calculations in relation to structural and electronic properties of few-layer graphitic-like AlN [4, 8] and GaN [9, 10] have herein been selected.

By following the predictive first-principles calculations based on both geometry optimization and verification of dynamic stability by phonon dispersion simulations, GaN (along with AlN and InN) has been found to form stable planar honeycomb monolayers; and stable planar bilayer AA´ stacking configurations (i.e., hexagons on top of each other with Ga atom over N atom and N atom over Ga atom).

Stability of planar graphitic-like structure of GaN has been disputed. Instead, for instance, its spontaneous reconstruction has been proposed into a structure that "remains hexagonal in plane but with covalent interlayer bonds that form alternating octagonal and square (8|4 Haeckelite) rings" [11]. As an example, a planar three-layer graphitic-like GaN has been found to represent a structure at a saddle point unlike the Haeckelite-type structure which has been found to represent a local minimum of the potential energy [11]. We note the calculated energy barrier of only 0.15 eV (~ 3.45 kcal mol$^{-1}$) between the graphitic-like and Haeckelite-type structures [11]. In principle, such a small barrier is not energetically significant. Even more, any energy barrier of less than 40–50 kcal mol$^{-1}$ is considered as relatively low and in the accessible energy range under MOCVD conditions of group III nitrides [12-14]. Hence, and mostly, deposition kinetics can dominate stabilization of various ultrathin structures of group III nitrides under MOCVD conditions in presence of a substrate and/or in confinement, even if they may not represent any local minimum of the potential energy.

In exploring different stacking configurations by predictive first-principles calculations, it has been reported that planar graphitic-like AlN can form energetically stable few-layer stacks with graphene [4]. Carrying out relaxations of the various stacking configurations, it has been established that certain deviation within the range of (0.01 – 0.1) Å of both the graphitic-like AlN and the graphene monolayers from their planarity may occur [4]. Such small deviation from planarity is perceived to have accommodated any long-range interlayer interactions between graphitic-like AlN and graphene; but it is unrelated to the common buckling parameter. In general, group III and group V atoms (which form the honeycomb structure of a III-V monolayer) may lie in different planes and therefore exhibit a certain buckling distance. A buckling distance in the range of 0.4 - 0.7 Å has been calculated in case of low-buckled honeycomb monolayer geometry which has been ascribed to common group III-V semiconductors such as InP, InAs, InSb, GaP and GaAs [5]. On the other side, the stable planar honeycomb monolayer structure has been ascribed to the group III nitrides (and all other binary compounds which contain B and C atoms) by first-principles calculations based on both geometry optimization and verification of dynamic stability by phonon dispersion simulations [5, 6]. As already discussed, by stacking planar honeycomb monolayers, it is possible to construct not only stable planar few-layer graphitic-like AlN [4, 8] and GaN [9, 10], but also multilayer 3D graphitic-like structures of AlN [8], and even newly predicted 3D periodic structures of GaN [10]. In the latter case [10], the buckled geometry with its associated onset of vertical Ga-N bonds between monolayers has been considered in detail. Noteworthy, the study in reference 10 points out to the "minute" energy difference between the planar bi-layer graphitic-like structure of GaN and other predicted buckled bi-layer structures. Therefore, the dominant effect of deposition kinetics in stabilization of various ultrathin structures of GaN may be reiterated again.

The stability of planar graphitic-like structure of GaN has been disputed in relation to the extent of passivation of any unsaturated states [15]. Thus, upon complete hydrogen passivation, monolayer as well as few-layer GaN have been investigated by predictive first-principles calculations and found to be energetically more stable in buckled geometry [15, 16].

In general, hydrogen composes most of the growth environment in the MOCVD of group III nitrides [17], including AlN [1], GaN[15], and InN [2] at the 2D limit. Other chemical species such as oxygen, though being credible only as trace impurities with less than 0.5 vol ppm in the gas stream in MOCVD growth environment, may also give rise to structural selectivity at atom-size-scale with implications for, e.g., material realization of 2D InO [18].

Herein, we report on monolayer GaN in a buckled geometry obtained in a framework of MOCVD processes, which so far have been shown to excel in the formation of graphitic-like AlN and InN confined at graphene/SiC interface [1, 2], and therefore probing a significant part of the landscape of group III nitrides at the 2D limit. Based on our experiments, we expose discontinuity in the anticipated stacking sequence of GaN monolayers which has been accounted for by formation of Ga-O bonds. On one side, the formation of Ga-O bonds may present implications for attaining a few-layer graphitic-like GaN. On the other side, the formation of Ga-O bonds acquires importance in instigating chemical-species-specific structural selectivity in confinement at atom-size scale.

## 2. Experimental

The MOCVD processes were performed in a horizontal-type hot-wall MOCVD reactor (GR508GFR Aixtron) which has been designed for the research and development of group III nitrides [17, 19]. Epitaxial graphene on a nominally on-axis 4H-SiC (0001) was employed with a possibility to decouple the zero-layer graphene from SiC by hydrogen intercalation [20]. The reactor was operated at a $H_2$ flow rate of 25 slm and a pressure of 200 mbar. Trimethylgallium, $(CH_3)_3Ga$, and ammonia, $NH_3$, precursors were employed at a flow rate of 0.442 sccm and 2 slm, respectively. The scheme of precursor delivery involved 3 cycles in total, each of about 3 minutes and consisting of alternating flows of $(CH_3)_3Ga$ with $NH_3$ and followed by an extra time of 10 minutes of their joint delivery. The flow of $NH_3$ and $H_2$ continued during the cooling down stage of the overall MOCVD process. The deposition of GaN was done at two different temperatures, 700°C and 1100°C.

The morphology of the samples was evaluated by tapping mode Atomic Force Microscopy (AFM) using Si probes and a DI3100 equipment with a Nanoscope V controller. Conductive atomic force microscopy (C-AFM) was further employed to probe vertical current injection through the graphene/SiC interface. Current mapping was carried out using Pt-coated Si tips with 5 nm curvature radius.

Transmission electron microscopy (TEM) was performed to define both structural and chemical details of the ultrathin films. TEM lamella were cut by focused ion beam (FIB) in which the energy of Ga ions was reduced to 2 keV in a dual beam FEI SCIOS2 equipment and initial observations carried out in a FEI THEMIS 200 microscope equipped with an image corrector at 200 keV. Scanning transmission electron microscopy (S/TEM) was carried out in a probe corrected JEOL ARM 200F microscope. High Angle Annular Dark Field (HAADF) S/TEM, with collection angles from 68 to 280 mrad was used to acquire *Z*-contrast images (which show an intensity dependence proportional to atomic number as $Z^2$) highlighting the presence of atomic Ga planes intercalated between bulk SiC and graphene, and to make direct measurements of interlayer distances at Ångström scale [21]. Annular Bright Field (ABF) S/TEM method was applied by collecting the electrons within a ring-shaped circumference from 11 to 24 mrad cutting the central part of the bright field disk out. The ABF method enables to effectively visualize atomic columns composed simultaneously of heavy and light atoms [22]. Actually, electrons of the beam scattered at higher angles, due to the interaction with the heavy atomic columns (high-*Z*), fall outside of the ABF detector with respect to the lighter ones (lower-*Z*). The heavy atoms are imaged as dark spots thus allowing even the lightest atoms, otherwise showing very weak contrast, to be visualized. Overall, the ABF method permits to measure the buckling distance which is not achievable with other characterization techniques.

Chemical analysis was performed with a TEM post-column Gatan Quantum ER Electron Energy Loss Spectrometer (EELS) with energy resolution of 0.3 eV operated in dual EELS mode which allows to simultaneously acquire spectra in the zero loss and high energy loss regions. Spectrum Imaging (SI) mode, by combining energy dispersive X-ray spectroscopy (EDS) and Electron Energy Loss Spectroscopy (EELS) pixel by pixel on the S/TEM image, was also used.

## 3. Results and discussion

The overall intercalation process into graphene/SiC interface under MOCVD conditions obeys several key features. Since the graphene/SiC interface is obtained in a high-temperature sublimation process, it is initially characterized by a high energy. It is a result to the partial bonding of the zero-layer graphene to the outmost Si atoms of the SiC surface and the prevailing large density of unsaturated Si dangling bonds. In general, the graphene/SiC interface may achieve surfaces of reduced energy via the process of hydrogen intercalation and associated regular termination of Si dangling bonds by hydrogen atoms. Such an enclosed space of surfaces of reduced energy enables lateral migration of (metal) atoms across large areas and synthesis of structures with atom-size thickness. This approach has been first-time demonstrated in achieving ultrathin 2D GaN structure of rhombohedral symmetry confined between graphene and SiC [15]. By extending this approach, 2D graphitic-like AlN [1] and InN [2] have also been achieved in MOCVD processes and with a very high coverage of the SiC surface [2].

Alternatively, in a typical intercalation approach - which involves deposition of metal atoms on graphene followed by annealing in an ultra-high vacuum system in the absence of hydrogen (see for example ref. 23 about intercalated In) - the high energy of the zero-layer graphene/SiC interface can act as a thermodynamic force for intercalation of metal atoms beneath the zero-layer graphene and minimization of the interfacial energy [24, 25]. The observed "spontaneous" intercalation of Ga (along with In) atoms reported in ref. 24 is a result to a plasma-assisted molecular beam epitaxy growth of GaN on graphene on SiC, whereas the intercalation of In and Ga atoms (along with Sn atoms) reported in ref. 25 is a result to a vaporization of metallic precursors situated in a crucible underneath the epitaxial graphene. A common feature in these intercalation processes [23-25] is the availability of free atoms of In and Ga.

The MOCVD processes typically involve trimethyl-based metalorganic precursors such as $(CH_3)_3Al$, $(CH_3)_3Ga$, and $(CH_3)_3In$ which exhibit strong metal-carbon bonding as expressed by the value of Gibbs free energy of formation and cohesive energy per atom [26]. Dissociative adsorption of these precursors advances upon their surface reactions with graphene having implication for, e.g., delivery of individual metal atoms to graphene surface. The nature of surface reactions of the most strongly bonded trimethylaluminum $(CH_3)_3Al$ precursor with self-standing graphene has been identified by purposefully developed ab-initio molecular dynamics simulations [27].

As described in the experimental part, the precursors, $(CH_3)_3Ga$ and $NH_3$, have been fed into the MOCVD reactor by following a discerned scheme at two different temperatures of 700 and 1100ºC. The temperature of 700°C has already proved successful for the intercalation of AlN [1] and InN [2] at the graphene/SiC interface by MOCVD in the same reactor; while a temperature in the vicinity of 1100 °C is typically used for epitaxy of GaN on SiC.

Initially, the samples have been characterized by conductive AFM (C-AFM). Particularly, current maps by C-AFM can provide local information on the electronic transport across graphene/SiC heterojunction with a lateral resolution corresponding to the effective contact size of the metallic tip (curvature radius of ~5 nm). A schematic of the employed experimental set-up is reported in **Figure 1**a. Two morphological images of the epitaxial graphene on SiC after intercalation at 700° C and 1100°C are shown in **Figure 1**b and **1**c, respectively. Furthermore, the corresponding current maps measured on the two sample regions are reported in **Figure 1**d and **1**e. Surface topography is very similar for the two samples, and it exhibits the typical step-bunching of 4H-SiC substrate originating from the high temperature graphene growth. On the other hand, significant differences can be observed between the two current maps, indicating a very different homogeneity of the intercalation at graphene/SiC interface at the two MOCVD temperatures. As recently demonstrated also in the case of intercalated InN [2], the presence of an ultrathin barrier layer causes a reduction of the injected current with respect to non-intercalated graphene/SiC areas. While uniformly intercalated areas (low current contrast) separated by non-intercalated ones (high current contrast) can be observed in the sample subjected to intercalation at 700°C (**Figure 1**d), the 1100°C MOCVD process results in the formation of small size intercalated patches (see **Figure 1**e) surrounded by non-intercalated regions. Finally, the percentage of intercalated area has been quantitatively estimated from the histogram of current distribution extracted from the two current maps, showing 66% intercalation after the 700°C MOCVD and only 3% intercalation after the 1100°C process.

The heating of the graphene/SiC in hydrogen to the higher temperature of 1100°C, which precedes the feeding with the $(CH_3)_3$Ga precursor, is expected to have resulted in a more complete decoupling of the zero-layer graphene from the Si-terminated SiC surface and its transformation into a 1$^{st}$ layer graphene. Consequently, the 1100°C process diminishes the areas which expose zero-layer graphene. Concurrently, the zero-layer graphene, due to bonding with underlying Si atoms of the SiC substrate, is more reactive and we have found that it facilitates, for example, the $(CH_3)_3$In/graphene reaction dynamics for supply of In atoms to the graphene surface [28]. We therefore speculate that the less percentage of Ga intercalation after the 1100°C MOCVD process may be a result to a less efficient dissociative adsorption of the $(CH_3)_3$Ga precursor.

*Figure 1.* *(a) Schematic of the C-AFM setup for vertical current mapping through the intercalated and non-intercalated interfaces. Surface morphology of epitaxial graphene on 4H-SiC after intercalation at 700°C (b) and 1100°C (c). Corresponding current maps on the same sample areas ((d) and (e)). Histograms of current distributions on the two samples ((f), (g)), from which the percentage of intercalated areas was estimated.*

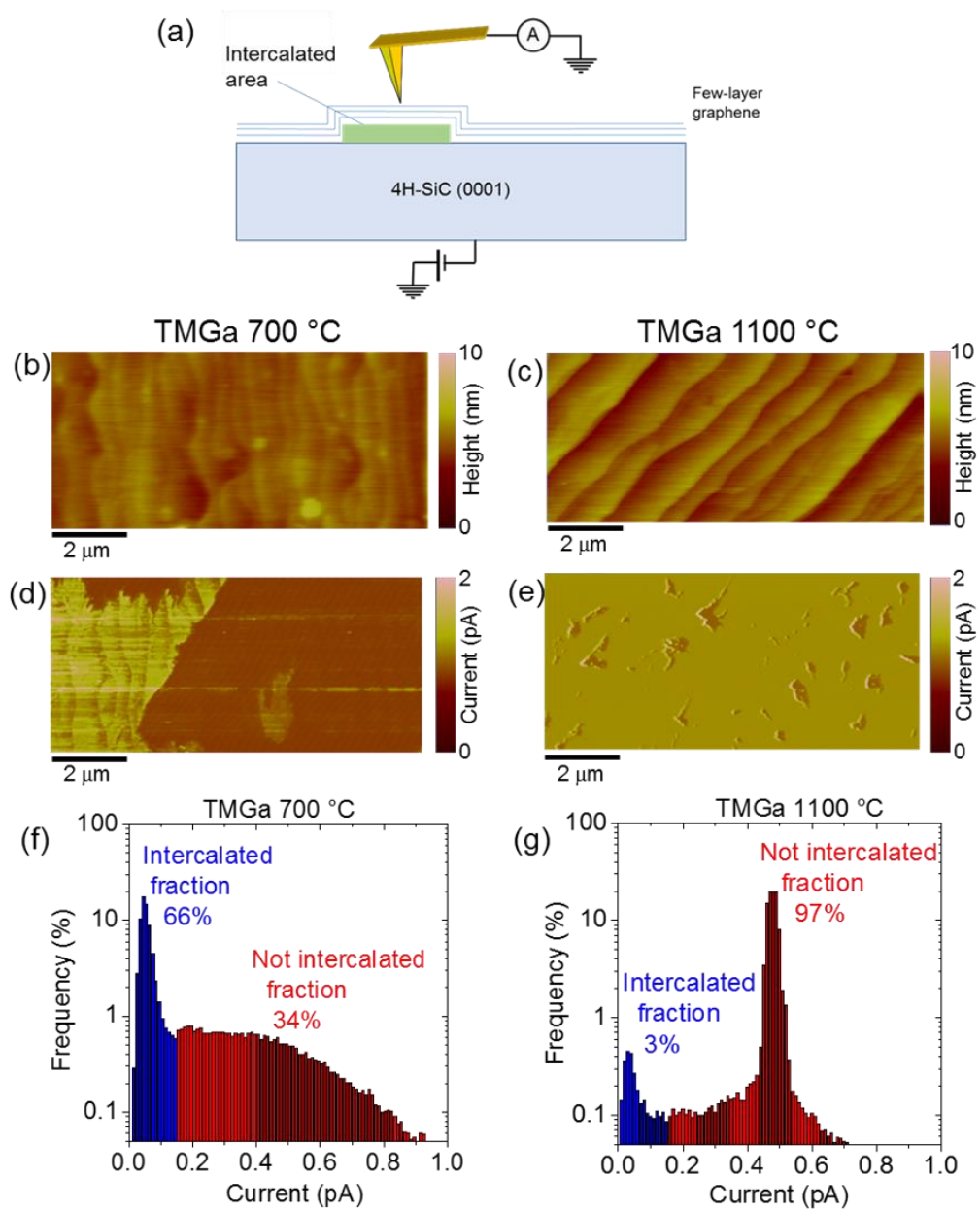

HAADF S/TEM imaging has confirmed intercalation by the presence of very bright contrast at the graphene/SiC interface which is related to the presence of higher atomic number Ga atoms (Z-contrast imaging). As a confirmation of this, **Figure 2**a and **2**c report in a comparative way HAADF S/TEM imaging of intercalated and non-intercalated graphene/SiC interfaces at 700ºC, respectively. Furthermore, for both micrographs, a linear intensity profile was extracted after integration over the area in the depicted boxes and used to measure the interplanar atomic distances, **Figures 2**b and **2**d. In this way, the characteristic interplanar distance of about 2.5 Å in bulk SiC, and 3.3 Å in bilayer graphene, respectively, could be confirmed, **Figure 2**d. The first carbon layer is displaced from the underlying SiC substrate at 2.8 Å, **Figure 2**d, and therefore indicative for its incomplete detachment [21] for which a displacement of about 3.6-3.8 Å is typically expected [29]. The spacing of about 2.8-2.9 Å between the intercalated Ga layers beneath a single layer of graphene in this case, is accordingly denoted in **Figure 2**c.

***Figure 2.*** *HAADF image and linear intensity profile, integrated over the area in the box, of intercalated (a, b) and non-intercalated (c, d) graphene/SiC interface. Images are acquired near the [11$\bar{2}$0] zone axis. Arrows in the boxes indicate the x-axis direction of intensity profiles. Distances reported in (b) and (d) are expressed in nm.*

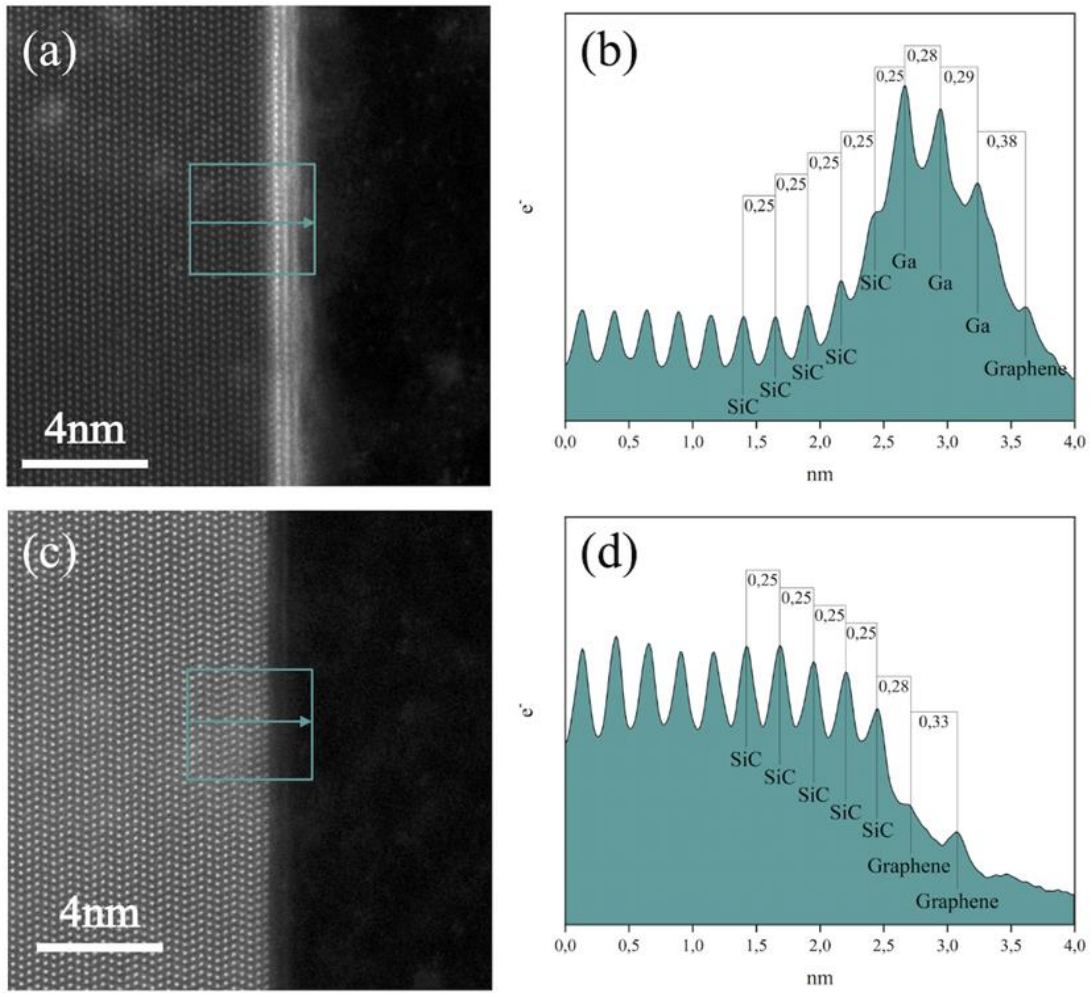

The intercalated layers at the graphene/SiC interface have been characterized in SI mode by combining EDX and EELS spectroscopy for compositional and electronic properties. SI/EDS was used to identify chemical species of interest, i.e., silicon, gallium, nitrogen, and oxygen, and to obtain bidimensional elemental maps as reported in **Figure 3**. The EDS analysis shows the contemporary presence of nitrogen and oxygen in correspondence to gallium layers. In particular, the nitrogen signal originates closer to the gallium-silicon interface while the oxygen signal comes from the outer gallium layers.

*Figure 3. ADF image of intercalated graphene/SiC acquired in SI mode and correlated EDS bidimensional elemental maps of silicon, nitrogen, gallium, and oxygen. Pixel size is 0.5 x 0.5 nm.*

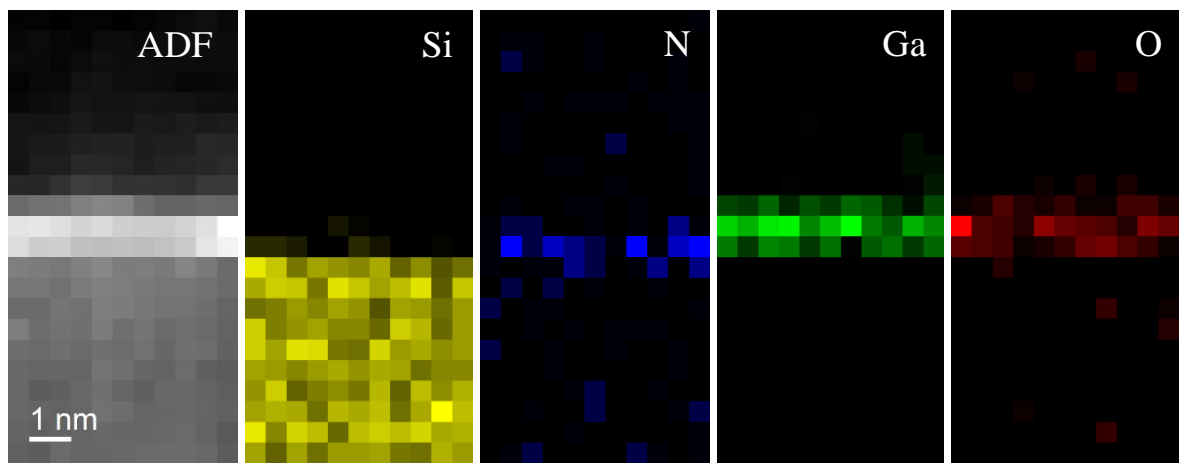

The SI/EELS investigation has provided information about the bonding and oxidation state of Ga across the intercalated structure (**Figure 4**). To increase the signal to noise ratio (SNR) of the EELS spectra, counts have been integrated within four rectangular boxes along the region of interest as displayed in the acquired ADF image (**Figure 4**a), so that each one encloses the area between two adjacent atomic layers (e.g., the first box between the last atomic layer of silicon and the first of gallium, the second box between the first two atomic layers of gallium), we may safely assume that electronic distribution along the integration area is the same. The four integrated spectra are reported in **Figure 4**b and from their study we obtain the following results. Spectrum 1, interface SiC/GaN, shows the presence of a nitrogen K-edge peak starting at 400 eV. In spectra 2 and 3, interface GaN/GaN, nitrogen signal significantly drops until it disappears in spectrum 4, interface GaN/graphene. Spectrum 1 also shows the presence of an oxygen K-edge peak near 530 eV whose intensity increases in spectrum 2, remaining constant over spectra 3 and 4.

***Figure 4** (a) ADF image of intercalated graphene/SiC. (b) Plots of EELS spectra acquired in SI mode in the high loss region of nitrogen and oxygen K-edge peaks, integrated over the corresponding area within the four different boxes. Pixel size is set at 0.05 x 0.05 nm. Amorphous carbon (a-C) at the top was deposited during FIB preparation of TEM lamella for surface protection during the milling.*

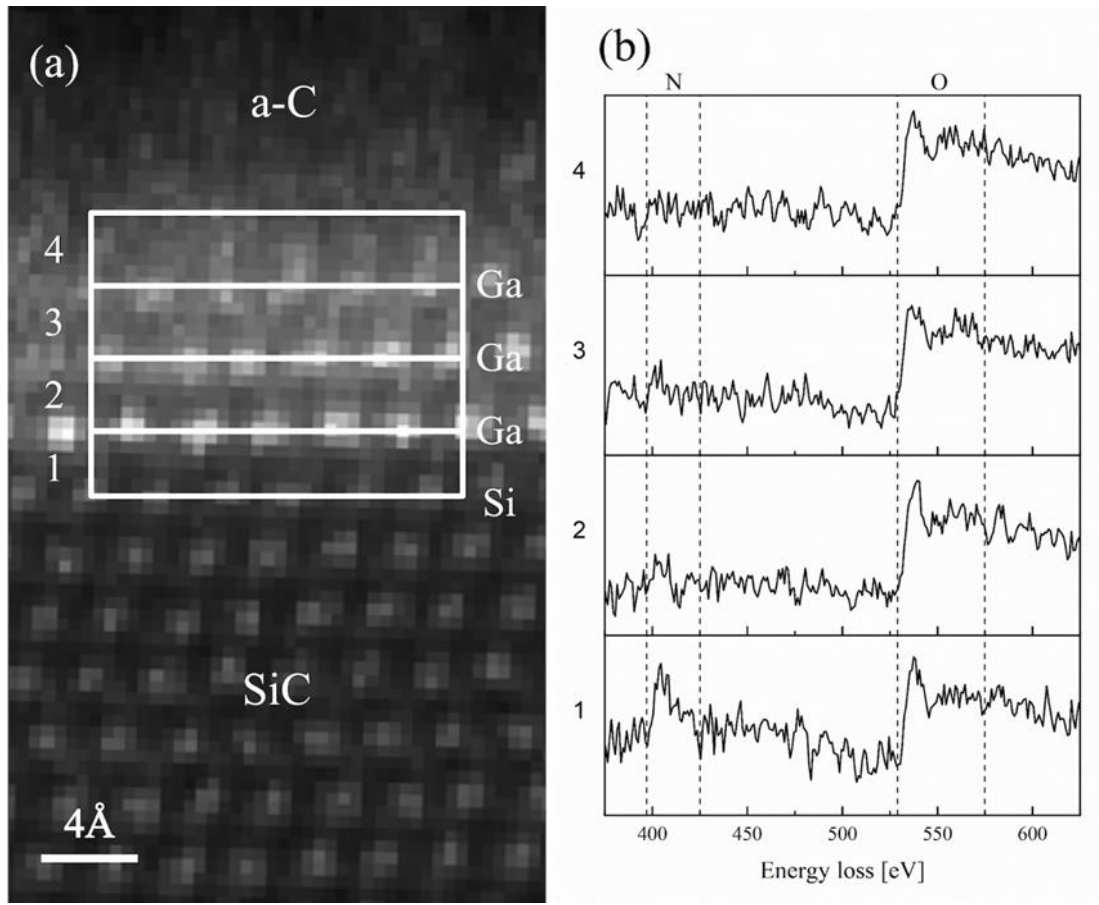

Remarkably, the EELS analysis allowed to detect a clear signal from a single atomic layer of nitrogen in spectrum 1 indicating the formation of a GaN monolayer at the immediate interface with the underlying SiC substrate. The oxygen peak in spectrum 1 could be related to a minor oxidation and/or signal delocalization from the upper layer. In spectrum 2 and 3, weaker nitrogen signals along with marked oxygen peaks indicate the oxidation of Ga which occurs in the two outer atomic layers.

SI/EELS spectra have also been acquired in bulk 4H-SiC and amorphous carbon deposited during FIB preparation of the TEM lamella to confirm that the recorded EELS oxygen signals across the intercalated structure have at most a negligible contribution from surface oxidation of the specimen (**Figure S1**).

EELS results, therefore, confirm the successful intercalation of Ga and N between the SiC and graphene whereas further continuous replacement of nitrogen atoms by oxygen atoms becomes apparently relevant across the intercalated structure and indicative for formation of Ga-O bonds.  The EELS results were corroborated by ABF S/TEM imaging which allowed to investigate the atomic order in the intercalated structure by visualizing lighter elements like nitrogen and oxygen (**Figure 5**).

*Figure 5.* *ABF S/TEM image of the atomic order in the investigated intercalated structure. We note a stacking order fault in the SiC substrate at the interface with the GaN monolayer (further details in* ***Figure S2****).*

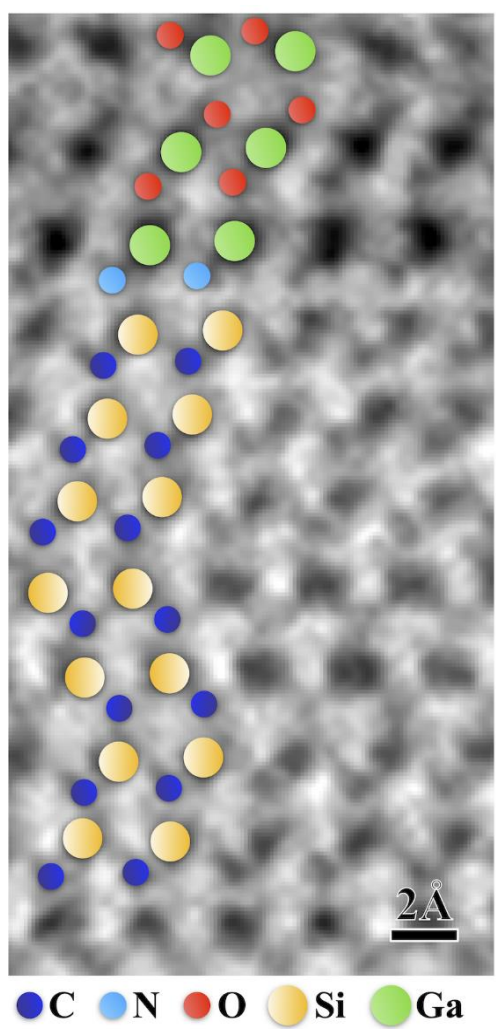

In the interpretation of the results presented above, we believe formation of a GaN monolayer in confinement at the graphene/SiC interface to be the primary process. Its buckled geometry can be indicated by the Ga and N atoms lying in different planes and having a certain buckling distance of 1.01 A ± 0.11 Å (our pixel size is 0.11 Å). Therefore, this value, extracted from experiment, complements existing knowledge about buckled honeycomb GaN monolayers which is solely based on predictive first-principles calculations. The buckling parameter of free-standing hydrogenated GaN monolayer has been calculated as $\Delta = 0.716$ Å [30]. Upon formation of a GaN monolayer, adsorption sites favorable for dissociation of, e.g., common trace impurity $O_2$ may become available on its Ga surface by overcoming an energy barrier of only 1.13 eV (~ 26.05 kcal mol$^{-1}$) which finds its reasoning in ab-initio calculations [31]. Oxygen species are inevitably present in MOCVD environment as common trace $H_2O$ and $O_2$ (< 0.5 vol ppm) in the gas stream. Thus, we speculate that availability of oxygen adatoms could change the bonding environment for (incoming) intercalated Ga atoms and cause a discontinuity in the anticipated stacking sequence of GaN monolayers. Instead, the Ga atoms in the topmost quintuple layer of the investigated intercalated structure have been found to have adopted an atomic order suggestive for ultrathin gallium oxide $Ga_2O_3$ [32]. The ultrathin structure of $Ga_2O_3$ is characterized by Ga atoms occupying sites of both, octahedral and tetrahedral coordination in O-Ga-O trilayer and Ga-O bilayer, respectively [32]. Upon the example of the established atomic order in the intercalated structure from **Figure 5**, we can validate a case of polarity inversion by simultaneous formation of Ga-N and Ga-O bonds. The simultaneous formation of Ga-N and Ga-O bonds may, in a general perspective, incur various other implications. It is tempting to relate to a possibility for stabilization in confinement of, commonly considered as a metastable, gallium oxynitride (GaON). It is known to exhibit statistical distribution between N and O atoms in the anion sites and vacancies in the Ga sites [33, 34].

## 4. Conclusions

Thus far, graphitic-like GaN has been predicted as theoretically possible but has not been experimentally reported. Beyond the predictions routinely achievable by first-principles calculations, and by means of MOCVD and advanced atomic resolution imaging and spectroscopy, we report GaN monolayer in a buckled geometry obtained in confinement at graphene/SiC interface. Discontinuity in the anticipated stacking sequence of graphitic-like GaN monolayers has been exposed and reasoned as a case of structural selectivity mediated by dissociative adsorption of oxygen on the GaN monolayer. Thus, the formation of Ga-O bonds acquires importance in instigating chemical-species-specific structural selectivity in confinement at atom-size scale.

**Corresponding Author**

*Anelia Kakanakova-Georgieva

Department of Physics, Chemistry and Biology (IFM), Linköping University, 581 83 Linköping, Sweden.

E-mail: anelia.kakanakova@liu.se

* Giuseppe Nicotra

Consiglio Nazionale delle Ricerche, Istituto per la Microelettronica e Microsistemi, Strada VIII, n. 5, Zona Industriale, I-95121, Catania, Italy.

E-mail: giuseppe.nicotra@cnr.it

**Author Contributions**

A.K.-G. conceived the research plan and developed the MOCVD processes. F.G. carried out the C-AFM measurements. B.P. took TEM images. STEM investigation and analytical measurements were carried out by G.N. and G.S. A.K.-G. conceptualized and wrote the paper with input from all authors. All authors discussed the results and commented on the manuscript.

**Acknowledgements**

The authors thank Dr. T. Iakimov and Prof. R. Yakimova for kindly providing samples of epitaxial graphene, and Dr. C. Bongiono and Dr. A.M. Mio for the useful discussions. This work was performed within the FLAG-ERA project GRIFONE. A.K.-G. acknowledges support from the Swedish Research Council (VR) through VR 2017-04071. F.G. acknowledges support from the Italian Ministry of Education and Research (MIUR) under the project EleGaNTe (PON ARS01_01007). B.P. thanks the support of TKP2021-NKTA-05 and VEKOP-2.3.3-15-2016-00002 of the European Structural and Investment Funds. F.G. and B.P. acknowledge the CNR-HAS 2019–2022 bilateral project GHOST II. The S/TEM was performed at BeyondNano CNR-IMM labs, which is supported by the Italian Ministry of Education and Research (MIUR) under project Beyond-Nano (PON a3_00363). Part of this work has received funding from the European Union's Horizon 2020 research and innovation programme under grant agreement No 823717 – ESTEEM3.